\documentstyle[12pt]{article}

\begin{document}
\bibliographystyle{unsrt}

\begin{flushright} UMD-PP-94-166

\today
\end{flushright}

\vspace{0mm}

\begin{center}
{\Large \bf   Automatically R-Conserving  Supersymmetric
SO(10) Models and Mixed Light Higgs Doublets }\\ [6mm]
\vspace{15mm}

{\bf Dae-Gyu Lee and R. N. Mohapatra}\\
{\it Department of Physics, University of Maryland\\  College Park,
Maryland 20742}\\ [4mm]

\vspace{20mm}

\end {center}

\begin{abstract}
In  automatic R-parity conserving supersymmetric (SUSY) SO(10) models,
 the simplest way to accomodate realistic fermion masses
 is to demand that the light Higgs doublets are linear combinations
 of the \{{\bf 10}\} and \{$\overline{{\bf 126}}$\} grand unified
Higgs representations. We study the realization
  of this mixed light Higgs property ( MLHP ) consistent
with doublet-triplet splitting in a
 minimal R-conserving SUSY SO(10) model. We then discuss
 predictions for neutrino masses and mixings in this model as well
as its implications for proton decay.

\end{abstract}

\newpage
\section{Introduction}
\hspace{8.8mm} Supersymmetric (SUSY) SO(10) model has a number of desirable
features that make it an ideal candidate to describe physics beyond
 the standard model in a grand unified framework. They are: \\
(a) unification of all fermions of each generation into a single spinor
representation\cite{SOT} restoring quark-lepton symmetry
 to weak interactions; \\
(b) a natural implementation of the see-saw mechanism\cite{SEE} for
understanding small neutrino masses, which in the minimal version of the model
are of the right order of magnitude to explain the solar neutrino puzzle via
the MSW\cite{MSW} mechanism; \\
(c) a simple mechanism for explaining the origin of matter
 starting with a zero baryon and lepton asymmetry of the universe
for temperature above the grand unification scale\cite{FY86}.

In this paper, we wish to discuss a subclass of SUSY SO(10) models
 which have another highly desirable feature: automatic R-parity
conservation which leads to natural conservation of
 baryon and lepton number symmetries
 prior to symmetry breaking. As is well-known, this property is not present
in the SUSY standard model nor in the SUSY SU(5) model\cite{SUF}, where extra
symmetries have to be imposed by hand to ensure R-parity invariance.
On the other hand, in the SO(10) model
 if all Higgs representations are chosen to have congruence
 number zero (such as {\bf 45}, {\bf 54}, {\bf 210}, etc.)
 and two (such as {\bf 10}, {\bf 120}, {\bf 126},
  $\overline{{\bf 126}}$, etc.), the
R-parity symmetry is automatic.
 Two possible minimal models with this property are given below: \\

\noindent {\bf \underline{Model A :}} The Higgs particles belong
 to representations \{{\bf 210}\}, \{{\bf 126}\} $\oplus$
 \{$\overline{{\bf 126}}$\} and \{{\bf 10}\}\cite{BM93}.
 The role of \{{\bf 210}\} is to break SO(10) down to
 SU(2)$_{L} \times$  SU(2)$_{R} \times$ SU(4)$_{C}$;
that of \{$\overline{{\bf 126}}$\} (denoted by $\overline{\Delta}$)
 is to break SU(2)$_{R} \times$ SU(4)$_{C}$ down
 to U(1)$_{Y} \times$ SU(3)$_{C}$ while at the same time giving
 heavy Majorana mass to the right-handed neutrino ($\nu_R$) to
implement the see-saw mechanism for neutrino masses;
 the role of \{{\bf 126}\} (denoted by $\Delta$) is to cancel
 the $\overline{\Delta}$
contribution to the D-term so that supersymmetry
 is maintained down to the weak scale; the role of \{{\bf 10}\}
 (denoted by $H$) is of course to break the SU(2)$_{L} \times$ U(1)$_{Y}$
 to U(1)$_{e.m.}$ and generate fermion masses.
The model is minimal in the sense that omitting any one of
these multiplets will leave extra undesirable
 local symmetries at low energies or break SUSY at GUT scale.

\noindent {\bf \underline{Model B :}} The Higgs particles
 belong to the multiplets \{{\bf 45}\} $\oplus$ \{{\bf 54}\}
 (denoted by $A$ and $S$ respectively), and \{{\bf 126}\}
 $\oplus$ \{$\overline{{\bf 126}}$\} $\oplus$ \{{\bf 10}\}.
 Apart from being more economical compared to model A, this
model  has another difference from model A. {\it i.e.},
 here SO(10) is broken down to
SU(2)$_{L} \times$  SU(2)$_{R} \times$ U(1)$_{B-L} \times$ SU(3)$_{C}$ before
breaking to $U(1)_Y\times SU(3)_C$.

It is worth pointing out that in general superstring models, the
 \{{\bf 126}\} $\oplus$ \{$\overline{{\bf 126}}$\} (or higher)
 SO(10) representations do not emerge below the Planck
 scale after compactification\cite{CCL}.
Therefore, several SO(10) models discussed in the literature
\cite{AD93} have only
\{{\bf 45}\}, \{{\bf 16}$_H$ + $\overline{{\bf 16}}_H$\}, and \{{\bf 10}\},
where \{{\bf 45}\}
$\oplus$ \{{\bf 16}$_H$\} $\oplus$ \{$\overline{{\bf 16}}_H$\}
are used to break SO(10) down to the
standard model. In these models,
 the vacuum expectation values(VEV's) of
 \{{\bf 16}$_H$\} $\oplus$ \{$\overline{{\bf
16}}_H$\} lead to R-parity breaking terms at low energies.
For instance, a term of the form $\Psi \Gamma_a \Psi \Psi
 \Gamma_a \Psi_H /M_{pl}$ ( where $\Psi$ denotes matter spinor
and $\Psi_H$ denotes Higgs spinor ) will
lead to  B-violating terms of the form
  $<\tilde{\nu_H^c}>(u^c d^c d^c + Q L d^c)/M_{pl}$,
which can lead to catastrophic baryon number violation\footnote{ The
existence of the $u^cd^cd^c$ terms was pointed out to R. N. M. by A. Farragi}.
One must invoke additional symmetries to prevent the R-non-invariant terms.
It may very well be that such symmetries emerge
 from superstring compactification. But this remains to be demonstrated.
Moreover, with the above minimal set, it is impossible to break the
SO(10) symmetry down to the standard model without including large
Planck scale induced non-renormalizable terms in the superpotential.

Let us now discuss the question of fermion masses in the R-parity conserving
SO(10) models. It is well known\cite{Chano} that if only a complex {\bf 10}
Higgs representation contributes to fermion masses, then one gets the
undesirable mass relations between leptons and quarks e.g. $m_d/m_s =
m_e/m_{\mu}$, which is a factor of 10 off compared to observations. One
way to correct it is to have the Higgs doublets contained in the
$\overline{\bf 126}$ contribute to the fermion masses through their
direct dimension 3 couplings. However, if we naively added a separate
$\overline{\bf 126}$, for this purpose, there would be an extra pair of
Higgs doublets at low energies which is unacceptable from the point of view of
gauge coupling unification\cite{GCU}. Moreover, the usual problem
of doublet-triplet splitting will also get worse. It was pointed out
in Ref.~\cite{BM93} that there is a simple solution to this problem:
the same $\overline{\bf 126}$, which contributes to the breaking of $B-L$
symmetry and to the see-saw mechanism, can have its weak doublets
acquire an induced  VEV
without leaving any extra light doublet and without any fine tuning
provided there is a coupling between the {\bf 10} and the $\overline{\bf 126}$
in the superpotential via other Higgs multiplets that only have VEV's of
order of the GUT scale. (Such couplings can arise, for instance, if there is
a {\bf 210} multiplet in the theory.) The induced VEV then will correct the
bad charged fermion mass relations.

This property of inducing VEV for {\bf (2,2,15)} of
 $\overline{\Delta}$ can also be stated in
another way. When SO(10) breaks down to the
standard model at M$_U$, there are two
light doublets (say $\phi_u$ and $\phi_d$).
 To obtain realistic quark lepton mass relations,
they  must be a linear combination of the doublets in
 \{{\bf 10}\} and \{$\overline{{\bf 126}}$\}. In the rest of the
paper, we will call this ``mixed light Higgs property"(MLHP).
It is simple to see that models with MLHP do not have the
property of Yukawa unification widely discussed in recent literature
\cite{Yukawa} but strict Yukawa unification is anyway not realistic
for the first and second generation.
This  property, MLHP does impose non-trivial constraints on model building.
For instance,  maintaining MLHP while implementing
the doublet-triplet splitting (DTS) is rather nontrivial in general.
 In the Dimopoulos-Wilczek scheme\cite{DW82} for DTS,
an example was constructed\cite{LM3}, where this property emerges
consistent with some discrete symmetries. In fact, it was
 suggested in Ref.~\cite{LM3} that any new DTS scheme must have
 this property. In this paper, we will demonstrate a simple
 SO(10) model which has this good light Higgs property
 and study the consequences of this general class of SO(10) models
with the additional minimality criterion.

This paper has been organized as follows: in Sec.~2, we argue in favour of
 the necessity of the MLHP in R-conserving SO(10) models;
 in Sec.~3, we discuss the superpotential of the model B and
 introduce the doublet-triplet splitting mechanism with MLHP; in Sec.~4,
 we study the quark and lepton mass matrices and discuss its
 implications for neutrino masses and mixings; in Sec.~5, we
 discuss the implications for proton decay; Sec.~6 is devoted
to an R-conserving SO(10) model where the Higgsino mediated
 contributions to proton decay is naturally suppressed
  while maintaining MLHP; in Sec.~7, we summarize our results and conclude;
in an appendix, we discuss the minimum of the Higgs potential
consistent with the desired symmetry breaking pattern.

\section{Mixed versus Pure Light Higgs}
\hspace{8.8mm} In this section, we explore to what extent, it is
 an absolute necessity to have mixed light Higgs doublets
in an automatically R-parity conserving SO(10) model. {\it i.e.}, could
 the light Higgs below GUT scale be purely light Higgs (PLH) arising
 solely from the complex \{{\bf 10}\}~? It is obvious that
 if non-renormalizable Planck induced terms are not included in the
 superpotential, then  PLH scheme will not work since it will lead
 to the bad fermion mass relation $m_d/m_s=m_e/m_{\mu}$ already mentioned.
 However, once the non-renormalizable Planck induced terms are included,
 the result is not obvious. To see what happens, let us assume
 first that the SO(10) is broken down to the standard model by \{{\bf 210}\}
 Higgs, via VEV's for the components ({\bf 1,1,1}) $\oplus$ ({\bf 1,1,15})
 $\oplus$ ({\bf 1,3,15}). The possible non-renormalizable terms
involving the matter superfields are of the form: $\Psi \Psi \Phi H$,
$\Psi \Psi \Phi^2 H$, etc. It is then easy to verify that fermion mass
 matrices will have the general form:
\begin{eqnarray}
M_u &=& (h +h^{'} +f +f^{'}) \kappa_u  , \nonumber \\
M_d &=& (h +h^{'} -f -f^{'}) \kappa_d , \nonumber \\
M_l &=& (h +h^{'} +3f +3f^{'}) \kappa_d  . \label{EXRTA1}
\end{eqnarray}
Here, $f$, $f^{'}$, and $h^{'}$ are of order $\sqrt{8 \pi} M_U/M_{pl}$;
 $f$ is symmetric coming from an effective $\overline{{\bf 126}}$
 operator; $f^{'}$ and $h^{'}$ are antisymmetric coming
 from effective {\bf 120} operators. One then has the relation
\begin{eqnarray}
M_l = 2r M_u -  M_d . \label{EXRTA2}
\end{eqnarray}
Taking trace of both sides of this equation, we get $r\simeq m_b/m_t$ or
 zero. The right hand side of Eq.2 is completely determined by
the known values of the quark masses and CKM angles. These values have
to be extrapolated to the GUT scale in order to test the sumrule in eq.2.
The details of this extrapolation procedure is described in Sec.4 and is
applied to the model B. Here we  use those extrapolated values of the
parameters to study the validity of Eq.2. We find that generically, the
electron mass comes out a factor of five or so too large
compared to the observed value with no free parameter left
to adjust.The muon mass is also in disagreement with the known value
by about a factor of 1.5.
 Therefore, we feel that it is highly unlikely that a pure light
Higgs possibility in the sense defined in this paper would be realistic.

Similar arguments apply if the GUT symmetry is broken by a combination of
\{{\bf 45}\} $\oplus$ \{{\bf 54}\}. Thus, we feel that the mixed light
Higgs doublet property provides a better chance to get a realistic fermion
 spectrum in a minimal R-conserving SO(10) model. In the
 next section, we give an example of an explicit model where
MLHP is realized and proceed to discuss its implications in
subsequent sections.

\section{Mixed Light Higgs in Model B}
\hspace{8.8mm} In this section, we present a simple minimal SO(10) model,
where the light Higgs doublets have the desirable property (MLHP) described
in the introduction. As already mentioned , the key step
is to have a term in the superpotential that couples the {\bf 10} Higgs
with the $\overline{\bf 126}$ Higg(s) via Higgs fields that have VEV's of
order of the GUT scale. If we do not allow for Planck scale induced
non-renormalizable terms, then we need a {\bf 210} Higgs to achieve this
goal; but as shown in\cite{LM3}, achieving this together with doublet-triplet
splitting is not very simple, although we did manage to construct
an example which is technically natural. In this paper, we will take
the point of view that one should include  non-renormalizable
Planck scale induced terms and
  we will keep only the lowest order Planck induced terms.
As we will see this allows for the construction of a rather simple model
with MLHP.

As usual, we assign the fermions to the 16-dimensional spinor
representation of SO(10). We denote them by $\Psi_a$ (where
 $a=1,2,3$ stands for generations) and we use the following
\underline{minimal} set of Higgs bosons needed for
 complete symmetry breaking: \\
(i) \{{\bf 45}\} $\oplus$ \{$\overline{{\bf 54}}$\}
 (denoted by $A$ and $S$ respectively) to break SO(10)
 down to the left-right symmetric group
SU(2)$_{L} \times$  SU(2)$_{R} \times$ U(1)$_{B-L} \times$ SU(3)$_{C}$; \\
(ii) \{{\bf 126}\} $\oplus$ \{$\overline{{\bf 126}}$\}
 to break the SU(2)$_{L} \times$  U(1)$_{B-L}$ symmetry
 down to U(1)$_{Y}$ while keeping
supersymmetry in fact down to the M$_W$ scale; \\
(iii) A single \{{\bf 10}\} (denoted by $H$) to break the
SU(2)$_{L} \times$  U(1)$_{Y}$ down to U(1)$_{e.m.}$.

The superpotential of the model is chosen to consist of the following parts:
\begin{eqnarray}
W=W_f +W_s +W_p, \label{SP}
\end{eqnarray}
where
\begin{eqnarray}
W_f &=& h_{ab}\Psi_a \Psi_b H + f_{ab} \Psi_a \Psi_b \bar{\Delta},
 \label{Wf} \\
W_s &=& (\mu_H + \lambda S) H H + \mu_s S^2 + \lambda_s S^3 + \mu_A A^2
+\mu_{\Delta} \Delta \bar{\Delta} \nonumber \\
& &+\lambda_{\Delta} \Delta A \bar{\Delta}
+\lambda_{S} ( S \Delta \Delta +S \bar{\Delta} \bar{\Delta} ), \label{Ws} \\
W_p &=& { \sqrt{8 \pi} \lambda_p \over M_{pl} } \Delta A^2 H. \label{Wp}
\end{eqnarray}

As noted in Ref.~\cite{BM93}, if we show that the light
 doublets in the model are  linear combinations of the
doublets in \{{\bf 10}\} and \{$\overline{{\bf 126}}$\} multiplets,
 then $W_f$ can accommodate a realistic charged fermion
spectrum for all generations. This is, of course, intimately
connected with the question of the doublet-triplet splitting.

It is easy to see (see Appendix), using the above superpotential,
 that the vanishing of F-terms at the scale M$_U$ and $v_R$
 (which are equal to fit low energy LEP data\cite{AD93}),
 is guaranteed by the following choice of VEV's for the
 Higgs fields $S$, $A$, $\Delta$, and $\overline{\Delta}$:

\begin{eqnarray}
<S> &=& diag(1,1,1,1,1,1,-{3 \over 2},-{3 \over 2},-{3 \over 2},
-{3 \over 2}) M_U, \nonumber \\
<A> &=& i \tau_2 \otimes diag(b,b,b,c,c), \nonumber \\
<\Delta>_{(1,3,\overline{10})} &=& <\overline{\Delta}>_{(1,3,10)} = v_R .
\label{VEV}
\end{eqnarray}
Using this, we find that the mass matrix for the fermionic
 doublets in $H$, $\Delta$, and
$\overline{\Delta}$ can be written as:
\begin{eqnarray}
\begin{array}{cc}
 & \qquad \bar{\Delta}_u \qquad \qquad \ \ \Delta_u \ \ \  \qquad   H_u \\
\begin{array}{c}
 \bar{\Delta}_d \\ \Delta_d \\ H_d
\end{array}
& \left(
\begin{array}{ccc}
\begin{array}{c}
 0 \\ \mu_{\Delta}-\lambda_{\Delta}c \\ 0
\end{array}
&\begin{array}{c}
 \mu_{\Delta}+\lambda_{\Delta}c \\ 0 \\ \mu_2
\end{array}
&\begin{array}{c}
0 \\ \mu_1 \\ 0
\end{array}
\end{array}
\right)
\end{array} ,  \label{DM}
\end{eqnarray}
where we have fine tuned $\mu_{H}-(3/2)\lambda M_U \approx M_W$
 (assumed to be zero in writing Eq.~(\ref{DM})). In  Eq.~(\ref{DM}),
the rows and columns denote the down- and up-type doublets contained
 in $H$, $\Delta$, and $\overline{\Delta}$ respectively.
 The $\Delta A^2 H$ entries are induced by the
Planck scale corrections to $W_p$ leading
 to $\mu_i \approx \lambda_p { \sqrt{8 \pi} M_U^2 / M_{pl} } $ ( i= 1,2 )
 which is of order 10$^{-1} M_U$ to 10$^{-2.5}M_U$.
Note that the ({\bf 1,3,1}) VEV in $<A>$ makes the
 entry $\Delta_u H_d$ different from
$\Delta_d H_u$ ({\it i.e.,} $\mu_1 \neq \mu_2$).

It is easy to see from Eq.~(\ref{DM}), that the light
 Higgs doublets (denoted by $\phi_u$ and $\phi_d$) are given by
\begin{eqnarray}
\phi_u &=& cos \alpha H_u +sin \alpha \overline{\Delta}_u, \nonumber \\
\phi_d &=& cos \gamma H_d +sin \gamma \overline{\Delta}_d,  \label{DOU}
\end{eqnarray}
where $tan \alpha = \mu_1 /(\mu_{\Delta}-\lambda_{\Delta}c$) and
$tan \gamma = \mu_2 /(\mu_{\Delta}+\lambda_{\Delta}c$).
It is important to note that in general $\alpha \neq \gamma$.
We find this to be a rather simple way to get the light doublets
 with the correct group theoretical property at
low energies, for fermion masses.
Furthermore, we expect $\alpha$ and $\gamma$ to be much smaller than
one so that the departure from strict Yukawa unification \cite{Yukawa}is small.

The triplet mass matrix is a four-by-four matrix as follows:
\begin{eqnarray}
\begin{array}{cc}
 &\hspace*{10mm} H \hspace*{25mm} \Delta \hspace*{24mm} \bar{\Delta}
\hspace*{16mm} \bar{\Delta}_R  \\
\begin{array}{c}
 H \\ \bar{\Delta} \\ \Delta \\ \Delta_R
\end{array}
& \left(
\begin{array}{cccc}
\begin{array}{c}
2\mu_H + \lambda M_U\\ 0 \\ q_1(-b^2+\alpha^{\prime} c^2) \\ q_2 bc
\end{array}
&\begin{array}{c}
q_1(b^2+\alpha^{\prime} c^2)  \\ \mu_{\Delta}-\lambda_{\Delta}
 \beta^{\prime} b \\ 0 \\ 0
\end{array}
&\begin{array}{c}
0 \\ 0  \\ \mu_{\Delta}+\lambda_{\Delta} \beta^{\prime} b \\ 0
\end{array}
&\begin{array}{c}
0 \\ 0 \\ 0 \\ \mu_{\Delta}+\lambda_{\Delta}\gamma^{\prime} b
\end{array}
\end{array}
\right).
\end{array}   \label{TM}
\end{eqnarray}
In  Eq.~(\ref{TM}), the first three rows denote the anti-quark-type
 Higgsinos contained in ({\bf 1,1,6}) of $H$, $\overline{\Delta}$,
 $\Delta$, and the last row denotes the same in the ({\bf 1,3,}
 $\overline{{\bf 10}}$) of $\Delta$
respectively; similarly, the first three columns denote the quark-type
 Higgsinos contained in ({\bf 1,1,6}) of $H$, $\Delta$,
 $\overline{\Delta}$, and the last column
the same in ({\bf 1,3, 10}) of $\overline{\Delta}$.
And the primed symbols represent non-zero group theoretical factors and
$q_i$ are proportional to $\mu_i$.
It is certain that all eigenvalues are of order $M_U$. This
 solves the doublet-triplet splitting problem.

We want to point out that the specific form
of the superpotential $W_S + W_p$ can be
derived by requiring invariance under a $Z_2$ symmetry, under
 which $S$, $H$, and
$\Delta$ are even and $A$, $\overline{\Delta}$ are odd. This
 symmetry for instance forbids
the $\overline{\Delta} A^2 H$ term. Coming to the matter
 part of the superpotential, $W_f$, the $\Psi \Psi \overline{\Delta}$ term
is forbidden but there is an allowed Planck induced
 term $f^{'} \Psi \Psi A \overline{\Delta}$
which essentially plays the same role as the second term in $W_f$.
The effective coupling
in the mass matrix is then given by $f^{'} v_R \sqrt{8 \pi}/M_{pl}$.

Before turning to a discussion of the fermion masses in the model, we wish
 to emphasize two points:
first, in discussing the fermion masses in any
R-conserving SO(10) model, one must first ensure that the light
Higgs doublets have the correct MLHP property consistent with the
doublet-triplet splitting. The absence of color singlet Higgs doublets
in {\bf 45} representation makes our mechanism  a good starting point
for model building;
secondly, this property of the mixed light Higgs doublets is not trivial
 to ensure while keeping all
color triplet Higgsinos heavy. For instance, if
 instead of \{{\bf 45}\}, we used a \{{\bf 210}\}
to break SO(10),
the doublet mass matrix  becomes a four-by-four matrix since
 it includes the ({\bf 2,2,10}) submultiplet of \{{\bf 210}\} and
several elements in this matrix must be engineered
to zero value to attain the MLHP goal\cite{LM3}. Of course,
 one could double the number
of \{{\bf 126}\} $\oplus$ \{$\overline{{\bf 126}}$\} multiplets
 such that one set contributed to
the light Higgs doublets while the other breaks B-L symmetry
 and the two remain totally
separate. However, in this case,
one must worry about the possibility of unwanted
pseudo-Goldstone supermultiplets which spoil gauge unification.

\section{The fermion sector and predictions for the neutrinos:}
\hspace{8.8mm} Let us now turn to the fermion mass matrices in this model.
It was shown that in Ref.~\cite{BM93},
 the fermion mass matrices in this model are
characterized by 12 parameters, which can all evaluated
 given the six quark masses,
three charged lepton masses and three CKM angles.
 The neutrino masses and mixings are
then completely predicted. We repeat this discussion with two differences from
Ref.~\cite{BM93}. First, we take the effect of the
 superpartners on the running of the gauge
and Yukawa couplings. Second, we take the
 effect of the top quark Yukawa couplings
on the running of the masses\cite{NAC}.
 We will follow Naculich\cite{NAC} below.
 The low energy superpotential for the model is given by:

\begin{eqnarray}
W_0=h_u Q \phi_u u^c + h_d Q \phi_d d^c + h_e L \phi_d e^c +\mu \phi_u \phi_d,
\label{W0}
\end{eqnarray}
where $h_u$, $h_d$, and $h_e$ are three-by-three
 matrices expressible in terms of the
SO(10) coupling matrices $h$ and $f$ in Eq.~(\ref{Wf}) as follows:
\begin{eqnarray}
h_u &=& h cos \alpha + f sin \alpha, \label{hu} \\
h_d &=& h cos \gamma + f sin \gamma, \label{hd} \\
h_e &=& h cos \gamma -3 f sin \gamma. \label{he}
\end{eqnarray}

As emphasized earlier $\alpha \neq \gamma$ is required.
 Otherwise all CKM angles
vanish. Moreover, in a strict supergravity framework, where
supersymmetry breaking is
implemented in the hidden sector, at $\mu=M_{pl}$, we have
$m_{\phi_u}^2=m_{\phi_d}^2=m_{0}^2$. Their extrapolation down
to the electroweak scale
is governed predominantly\cite{ALS} by $h_{u,33}$ and $h_{d,33}$ .
Correct symmetry
breaking pattern ({\it i.e.,} $\tan \beta >1$)
 also requires that $\alpha$ and $\gamma$ be
different from each other. The electroweak
symmetry is then broken radiatively so that,
\begin{eqnarray}
<\phi_u^0> &=& v sin \beta, \nonumber \\
<\phi_d^0> &=& v cos \beta. \label{pud0}
\end{eqnarray}
The mass matrices at GUT scale can then be written as: ($\mu = M_U$)
\begin{eqnarray}
\overline{M}_u &=& (\overline{h} + \overline{f}) v, \nonumber \\
\overline{M}_d &=& (\overline{h} r_1 + \overline{f} r_2) v, \nonumber \\
\overline{M}_l &=& (\overline{h} r_1 -3 \overline{f} r_2) v, \label{Mudl}
\end{eqnarray}
where
\begin{eqnarray}
\overline{h} = h cos \alpha sin \beta; \overline{f} = f sin \alpha sin \beta;
r_1 = {cos \gamma \over cos \alpha} cot \beta; r_2 = {sin \gamma \over
 sin \alpha} cot\beta . \label{r12}
\end{eqnarray}
This is now in the same notation as in Ref.~\cite{BM93}.In order
 to evaluate the matrices
$\overline{h}$ and $\overline{f}$, $r_1$ and $r_2$, we
 use the following sum rule derived
in Ref.~\cite{BM93}, {\it i.e.,}
\begin{eqnarray}
\overline{M}_l={ 4 r_2 r_1 \over r_2 - r_1} \overline{M}_u -{3 r_2 + r_1
 \over r_2 - r_1}
\overline{M}_d. \label{LMF}
\end{eqnarray}

We then take Tr $\overline{M}_l$, Tr $\overline{M}_l^2$,
 and Tr $\overline{M}_l^3$ to
obtain $r_2$ and $r_1$\cite{LV1}.
The light neutrino mass matrix is given by the see-saw formula\cite{MSW} to
be\cite{BM93}
\begin{eqnarray}
M_{\nu}=-M_{\nu^D} M_{\nu^M}^{-1} M_{\nu^D}^T, \label{Mv}
\end{eqnarray}
where
\begin{eqnarray}
M_{\nu^D} &=& { 3 r_1 + r_2 \over r_2 - r_1} \overline{M}_u-{4 \over r_2 - r_1}
\overline{M}_d, \label{MvD} \\
M_{\nu^M} &=&- {1 \over R} \left[{ r_1  \over r_2 - r_1}
 \overline{M}_u-{1 \over r_2 - r_1}
\overline{M}_d \right], \label{MvD}
\end{eqnarray}
with
\begin{eqnarray}
R={v sin \alpha sin \beta \over v_R}. \nonumber
\end{eqnarray}

Note that this light neutrino mass matrix defined at $v_R$
needs to be extrapolated to the weak scale; but since $\nu_R$
 decouples below $v_R$, there are only some over all anomalous
 dimensions of the effective light-Majorana neutrino mass\cite{NMR}.
 This effect is small and we will ignore it.

In order to carry out the numerical analysis,
 we choose the following values for the running masses:
\begin{eqnarray}
m_u=5.1 \mbox{MeV};  \qquad \qquad m_c=1.27 \mbox{GeV};  \qquad  \qquad
m_t(m_t)=166 \mbox{GeV}; \nonumber \\
m_d=(8.9+ds) \mbox{MeV}; \qquad  m_s=(.175+ss) \mbox{GeV};  \qquad m_b=4.25
\mbox{GeV}; \nonumber \\
m_e=.51 \mbox{MeV}; \qquad  \qquad m_{\mu}=105.6 \mbox{MeV}; \qquad  \qquad
m_{\tau}=1.784 \mbox{GeV}. \label{INM}
\end{eqnarray}
The $m_t(m_t)$ is obtained by taking the CDF\cite{CDF}
 mean value of 174GeV for the $m_t$ pole. The symbols $ds$ and
 $ss$ are left free to be fixed by the sum rule in Eq.~(\ref{LMF})
 along with $r_1$ and $r_2$.
The CKM angles are parameterized in terms of $s_{12}$, $s_{23}$,
and $s_{13}$, with $s_{12}=-0.221$, $s_{23}=0.043$, and $s_{13}=0.0045$ as
 our choice corresponding to the mean values from experiments\cite{PDG}.

We then extrapolate all masses to the SUSY breaking scale\cite{NAC}.
 The extrapolation factors are defined as
 $\eta_i=m_i(m_i \mbox{\ or 1GeV})/m_i(\mu_{SUSY})$. They are
\begin{eqnarray}
\eta_{u}=2.17; \qquad \eta_{c}=1.89;
 \qquad \eta_{t}= 1;\qquad \eta_{d}=2.16; \nonumber \\
\eta_{s}=2.16; \qquad \eta_{b}=1.47;
 \qquad \eta_{e}=\eta_{\mu}=\eta_{\tau}=1.02.
\nonumber
\end{eqnarray}

In order to extrapolate from $\mu_{SUSY}$ to $M_U$,
 we need to know $tan \beta$. We follow Naculich\cite{NAC} and assume
 $tan \beta < 40$ so that effects of all Yukawa couplings except that
of the top quark can be ignored. In this limit, the top Yukawa coupling
 effect is accounted for by the factor $B_t=0.88647$.
The GUT scale values of the various masses (denoted with bars) are given by
\begin{eqnarray}
m_{u}=\overline{m}_{u} \eta_{u} A_{u} B_t^{3}; \qquad
m_{c}=\overline{m}_{c} \eta_{c} A_{u} B_t^{3}; \qquad
m_{t}=\overline{m}_{t} \eta_{t} A_{u} B_t^{6}; \nonumber \\
m_{d}=\overline{m}_{d} \eta_{d} A_{d} B_t^{0}; \qquad
m_{s}=\overline{m}_{s} \eta_{s} A_{d} B_t^{0}; \qquad
m_{b}=\overline{m}_{b} \eta_{b} A_{d} B_t^{1}; \nonumber \\
m_{e}=\overline{m}_{e} \eta_{e} A_{e} B_t^{0}; \qquad
m_{\mu}=\overline{m}_{\mu} \eta_{\mu} A_{e} B_t^{0}; \qquad
m_{\tau}=\overline{m}_{\tau} \eta_{\tau} A_{e} B_t^{0}, \label{MEG}
\end{eqnarray}
where $A$-factors are the contributions of the gauge groups to the
 extrapolation and are numerically given by
 (choosing $\mu_{SUSY}=170$ \mbox{GeV})
\begin{eqnarray}
A_u=3.21; \qquad A_d=3.13; \qquad A_e=1.48. \label{Aude}
\end{eqnarray}
Some of
the mixing angles are also extrapolated and one has (for $ij=$ 13 and 23 )
\begin{eqnarray}
s_{ij}=\overline{s}_{ij} B_t^{-1}. \label{SEG}
\end{eqnarray}

In order to predict neutrino masses, we need the value of the parameter
$R={v sin \alpha sin \beta / v_R}$.
The value of $sin \alpha$ and $sin \beta$ are arbitrary, whereas $v=246$ GeV
 and $v_R$ is fixed by unification of the gauge couplings. In the
 absence of the heavy particle threshold corrections, one
 has $v_R \approx M_U \approx 2 \times 10^{16}$ GeV\cite{NAC}; but
 as has been noted for the case of
 non-SUSY SO(10) models\cite{LM5}, the threshold corrections can easily
introduce uncertainty of a factor of 10$^{\pm 1}$ in $M_U$ and $v_R$. We will
 therefore assume that $v_R \approx 10^{15}$ to $10^{16}$ GeV in what follows.

In our input, the signs of the fermion masses can be arbitray. There are
 many posibilities. Below we give the results
 for the choice of signs for masses
 that satisfy all the constraints of the model: \\

\noindent {\bf \underline{Case I :}} All masses chosen positive.
 In this case, we find $ds=-1.581$, $ss=-0.03533$, $r_1=0.00952$,
 and $r_2=0.00479$. The Eq.~(\ref{LMF}) for this
choice of $\{r_1,r_2\}$ leads to the values of
$\{\overline{m}_{e},\overline{m}_{\mu},\overline{m}_{\tau}\}=\{
0.0003377,0.0695543,1.18177\}$ to be compared with extrapolated values:
\{ 0.0003378, 0.0695548, 0,1.18177\}.
The predictions for neutrino masses and mixing for this case are given by
\begin{eqnarray}
M_{\nu} &=& R  \{ -0.679348,43.2421,704.332\}, \nonumber \\
V_l &=& \left(  \begin{array} {ccc}
           0.999238    &-0.0381721  & 0.00819604  \\
            0.0379246  & 0.998875   &0.0284804  \\
           -0.00927398 &-0.0281479  &0.999561

	   \end{array} \right). \label{SOL1}
\end{eqnarray}

A natural value for $R \approx 10^{-14}$ to $10^{-13}$ (since
 $v sin \alpha sin \beta \approx 10^2$GeV) depending on whether $v_R$
is $M_U$ or $M_U/10$, we get
$m_{\nu_{\tau}} \approx 7 \times 10^{-3}$ eV to $7 \times 10^{-2}$,
 and $m_{\nu_{\mu}} \approx 4 \times 10^{-4}$eV to $4 \times 10^{-3}$eV.
 The $\theta_{e \tau}$ mixing angle in this case is rather small;
 but $sin^2 2\theta_{e \mu} \approx 5 \times 10^{-3}$, which is of right
 order of magnitude to resolve the solar neutrino puzzle via the
 MSW mechanism\cite{HL94}. \\

\noindent {\bf \underline{Case II :}} For the choice of all  signs
 for masses to be negative, we get a  consistent fit
 to all charged fermion masses for $r_1=0.00637635$, and $r_2=0.0187798$.
The predictions for the neutrino sector in this case are
\begin{eqnarray}
M_{\nu} &=& R  \{-0.910023, -21.4474, -547.607\}, \nonumber \\
V_l &=& \left(  \begin{array} {ccc}
             0.995493    &0.0921015   &-0.0225942  \\
            -0.0939442   &0.990303    &-0.102345   \\
             0.012949    &0.104006    & 0.994492
	   \end{array} \right). \label{SOL2}
\end{eqnarray}

In this case, both $\theta_{e \tau}$ and $\theta_{e \mu}$ mixing angles
 are outside the MSW two neutrino solution
 given by Hata and Langacker\cite{HL94}.
Therefore, if the solar neutrino deficit situation continues
to remain as it is now, this solution will be ruled out. \\

\noindent {\bf \underline{Case III :}} We have found another fit to the sum
 rule in Eq.~(\ref{LMF}), for the choice of masses, where $m_c,m_d,$
and $m_s <0$ whereas all the remaining masses are chosen positive.
 The values of $r_1$ and $r_2$ are: $r_1=0.0100082$ and $r_2=0.072028$.
The predictions for neutrino masses and lepton mixing for this case are
 given by
\begin{eqnarray}
M_{\nu} &=& R  \{0.0662987, 4.60097, -2041.97\}, \nonumber \\
V_l &=& \left(  \begin{array} {ccc}
            0.722321      & 0.684518    &0.0984244 \\
            0.69126       &-0.718838    &-0.0736991 \\
            0.0203028     &0.121271     &-0.992412
 	   \end{array} \right). \label{SOL3}
\end{eqnarray}

Again, here all (mass difference)$^2$ are outside the range of
 the small angle as well as the large angle MSW solutions
 to the solar neutrino puzzle. Again, this
 solution can be tested by the solar neutrino data. \\

\noindent {\bf \underline{Case IV :}} This case is obtained by changing
 the signs of masses in case III. This corresponds
 to the choices $r_1=0.0175744$ and $r_2=0.00699614$.
The predictions for neutrino masses and lepton mixing in this case are
\begin{eqnarray}
M_{\nu} &=& R  \{-0.460917, -30.2621, -663.422\}, \nonumber \\
V_l &=& \left(  \begin{array} {ccc}
            0.99973       &0.0226327    &0.00523126 \\
           -0.0227722     &0.999339     &0.0283497  \\
           -0.00458617    &-0.0284611   &0.999584
 	   \end{array} \right). \label{SOL4}
\end{eqnarray}

Here, again the mixing angles are outside the range
 required by the MSW analysis of the present solar neutrino data.

\noindent {\bf \underline{Case V :}} In this case,
all masses are chosen positive except the electron mass and we find
$ds=-1.581$, $ss=-0.03833$, $r_1=0.00955$, and $r_2=0.00500$.
The predictions for neutrino masses and mixing  are given by
\begin{eqnarray}
M_{\nu} &=& R  \{-0.945754, 42.8837, 714.726\}, \nonumber \\
V_l &=& \left(  \begin{array} {ccc}
      0.999542     &-0.029543  &0.00655202 \\
      0.0294036    & 0.999359  &0.0204401  \\
     -0.00715168   &-0.020238  &0.99977
	   \end{array} \right). \label{SOL5}
\end{eqnarray}

A natural value for $R \approx 10^{-14}$ to $10^{-13}$
 (since $v sin \alpha sin \beta \approx 10^2$GeV) depending
 on whether $v_R$ is $M_U$ or $M_U/10$, we get
$m_{\nu_{\tau}} \approx 7 \times 10^{-3}$ eV to $7 \times 10^{-2}$, and
$m_{\nu_{\mu}} \approx 4 \times 10^{-4}$eV to $4 \times 10^{-3}$eV.
 The $\theta_{e \mu}$ and $\theta_{e \tau}$ mixing angles in
this case are rather small for the MSW mechanism to work. \\

\noindent {\bf \underline{Case VI :}} If we choose all
 masses to be negative except  the electron mass, we get a fit
consistent with all charged fermion masses for
 $r_1=0.00656483$, and $r_2=0.0187798$.
The predictions for the neutrino sector in this case are
\begin{eqnarray}
M_{\nu} &=& R  \{-1.02584, -23.7002, -616.621\}, \nonumber \\
V_l &=& \left(  \begin{array} {ccc}
   0.993278 &0.112199  &-0.0284463 \\
  -0.114706 &0.98706   &-0.112052  \\
   0.015506 &0.114562  & 0.993295
   	   \end{array} \right). \label{SOL6}
\end{eqnarray}

In this case, both $\theta_{e \tau}$ and $\theta_{e \mu}$
 mixing angles are outside the MSW two neutrino solution \cite{HL94}.
Therefore, if the solar neutrino deficit situation continues
 to remain as it is now, this solution will also be ruled out.

If we set aside prejudices towards mixing angles coming
from solar neutrinos, then our $\nu_{\mu}$-$\nu_{\tau}$ mixing
angles are in the interesting ranges to be testable
 in the next generation of proposed acceleration
$\nu_{\mu}$-$\nu_{\tau}$  oscillation experiments.
In Fig.~1, we compare our predictions with the domains of
$\Delta m^2$ and $sin^2 2 \theta_{{\nu_\mu}{\nu_\tau}}$
angles to be explored in the proposed CERN and Fermilab experiments.

\section{Proton Decay}
\hspace{8.8mm} One of the key predictions of grand
 unified theories is the life-time of the
proton and its decay mode. In non-SUSY GUT models,
 the dominant decay of the proton
arises from the exchange of superheavy gauge bosons
 and the operators responsible for this have
dimension six. The primary decay mode is  $p \rightarrow e^+ \pi^0$.
On the other hand, in SUSY GUT models, in addition to the above
dim.-six operators, there also exist dim.-five operators, and
 in simple SUSY SU(5) or SUSY
SO(10) models, the latter graphs dominate.
 The resulting dominant decay mode is $p
\rightarrow \overline{\nu}_{\mu} K^+$, which
 can be used to distinguish between the SUSY
GUT theories from non-SUSY GUT ones.

Proton decay in SUSY SU(5) model has been extensively studied
\cite{AN93,HMY} and it
has been established that in this case,
 one requires the superheavy color-triplet Higgsino
($\tilde{H}_3$) mass $M_{\tilde{H}_3}\gg M_U$ in order
to be consistent with the existing
lower bounds on the
$\tau_{p \rightarrow \overline{\nu}_{\mu} K^+}$\cite{PDG}. We will show
below that in the SUSY SO(10) model dim.-five proton decay operators receive
contributions from two diagrams: one involving \{{\bf 10}\} Higgs and
the other involving
\{$\overline{{\bf 126}}$\} Higgs and these
graphs could interfere destructively, thereby reducing the
 effective $p$-decay amplitude.
This in turn can relax the constraints on color-triplet Higgsino masses.

The color-triplet Higgsinos that mediate proton decay
are part of a four-by-four matrix, given in Eq.~(\ref{TM}).
One can always find two four-by-four unitary matrices $V$ and
 $U$, for the triplet mass matrix Eq.~(\ref{TM}), such that
\begin{eqnarray}
(V^{\dagger})_{ik} (M_T)_{kl} U_{lj} = M_i \delta_{ij}. \nonumber
\end{eqnarray}
Then, the dim.-five operators at the GUT scale, which are to
 be turned into baryon-number
violating four-fermion interactions by gaugino- or
 Higgsino-dressing at the electroweak
scale, have the following common factor:
\begin{eqnarray}
C_{abcd}= C_0 \sum_{i=1}^4 {(h_{ab} U_{1i} + f_{ab} f_0 U_{3i})
 (h_{cd}V_{1i}^{*}+f_{cd}
f_0 V_{2i}^{*}) \over M_i}, \label{Cabcd}
\end{eqnarray}
where the early Latin indices are family ones and run over 1,2,3; $C_0$
and $f_0$ represent some over all factors (Clebsch-Gordan coefficients).

{}From the definition Eq.~(\ref{Mudl}), we obtain
\begin{eqnarray}
\overline{h} &=& {1 \over v} {r_2 \overline{M}_u - \overline{M}_d \over
 -r_1+r_2},  \nonumber  \\
\overline{f} &=& {1 \over v} {-r_1 \overline{M}_u +
 \overline{M}_d \over -r_1+r_2}. \nonumber
\end{eqnarray}
The first point to note is that since for all our solutions in Sec. 4
 $r_1$ + $r_2 \approx 5 \times 10^{-2}$ to $5 \times 10^{-3}$, the
proton decay amplitudes receive an extra enhancement factor $\approx 20$
 to 200 (depending on the cases) compared to minimal SU(5). However,
 unlike the SU(5) case, we have two separate contributions
 to the proton decay amplitude and we could hope to invoke parameters
 for which there is a cancellation. To see if this is possible,
 we choose the special case where $b^2= \alpha^{\prime} c^2$.
For convenience, we define
\begin{eqnarray}
\eta_{ab}={h_{ab} \over f_{ab}}.\nonumber
\end{eqnarray}
We, then, get
\begin{eqnarray}
C_{abcd}=C_0 h_{ab} f_{cd} \sum_{i=1}^3 {U_{1i}
  (\eta_{cd} V_{1i}^{*}+ f_0 V_{2i}^{*}) \over M_i}. \nonumber
\end{eqnarray}
(Note that one of the triplet pairs completely decouples from the proton
 decay amplitude.) We have $\eta_{11} \neq \eta_{22} \neq \eta_{12}$.
We, thus, have three equations involving the three mixing angles
 that characterize the matrices $V_{1i}$ and $V_{2i}$ and
 we could therefore expect a solution for which the proton
 decay is suppressed. It must however be pointed out that
 this does require a fine tuning of parameters.

\section{Further Suppression of Proton Decay Amplitude}
\hspace{8.8mm} In the previous section, we showed that
the minimal R-parity conserving
SO(10) model of Sec. 2 tends to predict higher strengths
 for the dim.-five proton decay
operators compared to minimal SUSY SU(5); however, unlike the minimal
SUSY SU(5) model, here there is the possibility of
cancellation if one allows fine-tuning among the mixings
 and masses for the color-triplet
Higgsinos. It is, however, worth emphasizing that
 the predictions for the neutrino sector and
realization of realistic quark-lepton masses are logically
independent of the proton life-time
predictions. The question, therefore, arises as to whether
 it is possible to suppress the
Higgsino-mediated proton decay amplitude while at the same
 time keeping the discussion
of light Higgs doublets with correct group theoretical
properties to give the mass matrix
structure of Sec. 4. A discussion of how to suppress such
 amplitudes in a different class of
SO(10) models has recently given in Ref.~\cite{BB93}. We
 present a different example that
has the addition property of ``mixed light Higgs doublet"
property (as defined in the
introduction).

We extend the model by addition of one more Higgs in the \{{\bf 10}\}
 representation
(denoted by $H_2$). Note that automatic R-conservation
 is maintained in the model. We
choose the superpotential to be
\begin{eqnarray}
W^{'} &=& \mu_2 H_2^2+\mu_{12} H_1 H_2 +\lambda_s S H_1 H_2 +\mu_s S^2
+\lambda_s^{'} S^3 +\mu_A A^2 \nonumber \\
& &+ \lambda_A S A^2 +\mu_{\Delta} \Delta \overline{\Delta} +
\lambda_{\Delta} \Delta A
\overline{\Delta} +{\lambda_p \over M_{pl}} \Delta A^2 H_1. \label{W'}
\end{eqnarray}
We require all $\mu \approx M_U$ and the
 fine-tuning condition $\mu_{12}+\lambda_s
(-3/2) M_U \approx M_W$. It is then clear that the light
 doublet mass matrix in Eq.~(\ref{DM}) is preserved.

In Table~I, we present a discrete
 $Z_3 $ symmetry under which all but
dimension two terms are invariant, providing a symmetry basis for this idea.
Due to the presence of dim.-two terms, this symmetry is softly broken,
 which in the supersymmetric context means that even after
 SUSY breaking terms are included no hard
dim.-four term would be generated with infinite coefficient
 that break the symmetry (thereby helping to maintain our  conclusion).
This symmetry has the implication that
 it allows the following matter couplings
 that will eventually lead to realistic fermion
masses and mixings if we assign $\Psi_a \Psi_b$ to be $\omega$
under the symmetry.
\begin{eqnarray}
W_m^{'}=h_{ab} \Psi_a \Psi_b H_1 + {f_{ab} \over M^2}
 \Psi_a \Psi_b A^2 \overline{\Delta},
\label{Wm'}
\end{eqnarray}
where $M=M_{pl}/\sqrt{8 \pi}$.
 The second term leads to both \{$\overline{{\bf 126}}$\},
\{{\bf 120}\} as well as \{{\bf 10}\} type couplings. The mass matrices are
 therefore less predictive than before. \\

\section{Summary and Conclusions}
\hspace{8.8mm} In summary, we have studied the question of how to get realistic
fermion masses in automatically R-parity conserving SUSY SO(10)
models. We have argued that an economical way to do this is to have the light
Higgs doublets contain a piece from the $\overline{\bf 126}$ Higgs multiplet
that is responsible for the breaking of the $B-L$ symmetry. Strict Yukawa
unification does not hold in these models\cite{Yukawa}.
 But, of course, it is well known  that models with strict
 Yukawa unification do not address the problem of second and
 first generation masses. We have given two examples of
such models and studied their predictions for neutrino masses and mixings
as well as proton decay. We find that all the models are testable by
the solar neutrino experiments as well as the proposed accelerator
experiments. In the proton decay sector, the situation is less predictive
than in the case of the minimal SU(5) model
 due to the presence of two different
contributions. We have also given a model in which the higgsino mediated
proton decay can indeed be arbitrarily suppressed.

\section*{Appendix}
\hspace{8.8mm} In this appendix, we will discuss the symmetry breaking
 of the SO(10) model down to the standard model for the choice of Higgs
 multiplets
$S$ \{{\bf 54}\}, $A$ \{{\bf 45}\},  $\Delta$ \{{\bf 126}\}
$\oplus$ $\overline{\Delta}$ \{$\overline{{\bf 126}}$\}.
The relevant part of the superpotential is given by $W_s$ in
Eq.~(\ref{Ws}). To this we add the Lagrange multiplier
 term $g$ $Tr S$ so that we can carry out the variation of all
 elements of the ten-by-ten symmetric matrix representing
\{{\bf 54}\} $\oplus$ \{{\bf 1}\}. We
will go to a basis where $S$ is diagonal without loss of generality.
 Let us now write down the constraints implied by all
 F-terms being zero. We will look for solutions with
\begin{eqnarray}
<S> &=& diag(1,1,1,1,1,1,-{3 \over 2},-{3 \over 2},-{3 \over 2},-{3 \over 2})
 M_U, \nonumber \\
<A> &=& i \tau_2 \otimes diag(b,b,b,c,c), \nonumber \\
<\Delta>_{\nu^c \nu^c} &=& <\overline{\Delta}>_{\overline{\nu}^c
 \overline{\nu}^c} = v_R , \nonumber \\
<H> &=& 0.  \label{APPE1}
\end{eqnarray}
{}From the equation $F_s=0$,
\begin{eqnarray}
2 \mu_s M_U + 3 \lambda_s M_U^2 - \lambda_A b^2 + g=0, \label{APPE2} \\
2 \mu_s (-{3 \over 2}M_U) + 3 \lambda_s ({9 \over 4}M_U^2) -
\lambda_A c^2 + g=0. \label{APPE3}
\end{eqnarray}
Demanding that $Tr S=0$ determines $g$ as follows:
\begin{eqnarray}
g=\left[-{9 \over 2} \lambda_s M_U^2 + {1 \over 5} \lambda_A
 (3 b^2 +2 c^2) \right]. \label{APPE4}
\end{eqnarray}
The other constraints are from vanishing of $F_A$,
$F_{\Delta}$ and  $F_{\overline{\Delta}}$ respectively:
\begin{eqnarray}
-2 \mu_A b -2 \lambda_A b M_U + \lambda_{\Delta} x_0 d^2=0,  \label{APPE5} \\
-2 \mu_A c +3 \lambda_A c M_U + \lambda_{\Delta} d^2=0,  \label{APPE6} \\
 \mu_{\Delta} d + \lambda_{\Delta} (x_0 b + c) d=0,  \label{APPE7}
\end{eqnarray}
where $x_0$ is a Clebsch-Gordan (C-G) coefficient.

Using the above constraints, we find the doublet Higgsino
matrix with quantum numbers ({\bf 2,1,1}) or ({\bf 2,-1,1})
 (under SU(2)$_L \times$ U(1)$_Y \times$ SU(3)$_C$) to be that
 given in Eq.~(\ref{DM}). As already mentioned,
 it contains only one pair of light doublets whose bosonic partners
 will be used to break the SU(2)$_L \times$ U(1)$_Y$.
For the color-triplet Higgsino matrix with
 quantum numbers ({\bf 1,-2/3,3}) or ({\bf 1,2/3,3$^*$}) to be
 that given in Eq.~(\ref{TM}). As already mentioned,
 it does not have zero eigenvalues.

The Goldstone modes are contained in the following mixings: \\
\noindent  ({\bf 2,1/3,3}), ({\bf 2,-1/3,3$^*$}); $S_{2,2,6}$,
$A_{2,2,6}$, $\Delta_{2,2,15}$, $\overline{\Delta}_{2,2,15}$ \\
\noindent  ({\bf 2,-5/3,3}), ({\bf 2,5/3,3$^*$}); $S_{2,2,6}$, $A_{2,2,6}$, \\
\noindent  ({\bf 1,4/3,3}); $A_{1,1,15}$, $\overline{\Delta}_{1,3,10}$ \\
\noindent  ({\bf 1,-4/3,3$^*$}); $A_{1,1,15}$, $\Delta_{1,3,\overline{10}}$ \\
\noindent  ({\bf 1,2,1}); $A_{1,3,1}$, $\Delta_{1,3,\overline{10}}$ \\
\noindent  ({\bf 1,-2,1}); $A_{1,3,1}$, $\overline{\Delta}_{1,3,10}$ \\
\noindent  ({\bf 1,0,1}); $S_{1,1,1}$, $A_{1,1,15}$,
  $A_{1,3,1}$, $\Delta_{1,3,\overline{10}}$, $\overline{\Delta}_{1,3,10}$.

Now, only two mixings are left. They are \\
\noindent  ({\bf 1,0,8}); $S_{1,1,20^{'}}$, $A_{1,1,15}$, \\
\noindent  ({\bf 3,0,1}); $S_{3,3,1}$, $A_{3,1,1}$.

We have exhausted all the submultiplets contained in $A$.
 The other submultiplets of $S$,  $\Delta$, and $\overline{\Delta}$
 cannot be mixed.
The remaining submultiplets in $S$ are \\
\noindent  ({\bf 1,-4/3,6}), ({\bf 1,4/3,6$^*$}); $S_{1,1,20^{'}}$, \\
\noindent  ({\bf 3,2,1}), ({\bf 3,-2,1}); $S_{3,3,1}$. \\
Their masses are given by
\begin{eqnarray}
|2 \mu_s + 6 \lambda_s c_i M_U |, \label{APPE9}
\end{eqnarray}
where $c_i$ are C-G coefficients.  We have checked that
 their masses are of order $M_U$, using Eqs.~(\ref{APPE2})-(\ref{APPE4}),

The remaining submultiplets are contained in
 $\Delta$ or $\overline{\Delta}$, and they are unmixed.
Their masses have the following form:
\begin{eqnarray}
|\mu_{\Delta}+\lambda_{\Delta}(x_i b + y_i c)|, \label{APPE8}
\end{eqnarray}
where $x_i$ and $y_i$ are C-G coefficients. To confirm that
 all the submultiplets are heavy, it is sufficient
 to show that $x_i$ in Eq.~(\ref{APPE8}) cannot be
$x_0$ in Eq.~(\ref{APPE7}) or that $y_i$ in Eq.~(\ref{APPE8}) cannot be 1.
{}From simple group theoretical consideration, $y_i=0$ for
 the submultiplets in  $\overline{\Delta}_{1,1,6}$, and
 $\overline{\Delta}_{3,1,\overline{10}}$. For the submultiplets
 which have the quantum numbers ({\bf 1,3,-2,1}), ({\bf 1,3,-2/3,3}),
and ({\bf 1,3,2/3,6}) (under SU(2)$_L \times$ SU(2)$_R \times$ U(1)$_{B-L}
 \times$ SU(3)$_C$), contained in $\overline{\Delta}_{1,3,10}$,
 the ratios\cite{L2} of $x_i$ are 1:1/3:1/3.
 The submultiplet with the quantum numbers ({\bf 1,-4,1}) in the
  ({\bf 1,3,-2,1}) has the value $y_i=-1$. Considering tensor indices
 of $\overline{\Delta}_{2,2,15}$ and the ({\bf 1,0,1}) in $A_{1,1,15}$,
we find that $x_i=0$ for the submultiplets in  $\overline{\Delta}_{2,2,15}$.
 The same is true for the corresponding submultiplets in $\Delta$. Thus
it is guaranteed by Eq.~(\ref{APPE7}) that all submultiplets whose masses
 are given by Eq.~(\ref{APPE8}) are superheavy.
Also, we have checked that except the Goldstone
 modes all submultiplets involved in the above mixings have superheavy masses.

\begin{center}
{\bf Acknowledgement}
\end{center}

This work has been supported by a grant from the National Science
Foundation.

\newpage
\section*{Table Caption}
Table I: $Z_3 $ symmetry quantum numbers for various fields. ($w=e^{i
2\pi/3}$.)

\vspace*{15mm}
\begin{center}
Table I  \\
\begin{tabular}{ |c|c| } \hline\hline
Fields    &$Z_3$             \\ \hline
$S$       &$\omega$       \\
$A$       &$\omega$      \\
$H_1$     &$\omega^2$    \\
$H_2$     &1             \\
$\Delta$  &$\omega^2$    \\
$\overline{\Delta}$  &1    \\  \hline\hline
\end{tabular}
\end{center}
\vspace*{25mm}

\section*{Figure Caption}
\vspace*{5mm}
\noindent Fig. 1:
This figure shows the present limits on $\nu_{\mu} \nu_{\tau}$ oscillation
parameters
($\Delta m^2$ and $sin^2 2 \theta$) and future possibilities on two proposed
experiments
CHORUS at CERN and P860A at Fermilab. The solid vertical lines are the
predictions of
the minimal SO(10) model
described in this paper for the six allowed parameter ranges,
denoted as cases I through VI in the text.

\end{document}